\documentclass[prl,superscriptaddress,twocolumn,preprintnumbers]{revtex4}

\usepackage{color}
\usepackage[active]{srcltx}
\usepackage{amsmath,amsfonts,amssymb,amsthm,amstext,amscd,eucal,srcltx}
\usepackage{epsfig,graphicx,bm}
\usepackage{epstopdf, epsf}
\usepackage{dcolumn}
\usepackage{hyperref}

\newcommand{\be}{\begin{equation}}
\newcommand{\ee}{\end{equation}}

\newcommand{\bse}{\begin{subequations}}
\newcommand{\ese}{\end{subequations}}
\newcommand{\bea}{\begin{eqnarray}}
\newcommand{\eea}{\end{eqnarray}}
\newcommand{\ba}{\begin{array}}
\newcommand{\ea}{\end{array}}
\newcommand{\bc}{\begin{center}}
\newcommand{\ec}{\end{center}}

\begin{document}
\preprint{IPM/P-2012/009} 
\vspace*{3mm}
\title{Kac-Moody instantons in space-time foam as an alternative solution to the black hole information paradox}

\author{Andrea Addazi}
\email{andrea.addazi@lngs.infn.it}
\affiliation{Center for Field Theory and Particle Physics \& Department of Physics, Fudan University, 200433 Shanghai, China}

\author{Pisin Chen}
\email{pisinchen@phys.ntu.edu.tw}
\affiliation{Leung Center for Cosmology and Particle Astrophysics, National Taiwan University, Taipei, Taiwan 10617}
\affiliation{Department of Physics, National Taiwan University, Taipei, Taiwan 10617}
\affiliation{Kavli Institute for Particle Astrophysics and Cosmology, SLAC National Accelerator Laboratory, Stanford University, Stanford, CA 94305, U.S.A.}

\author{Antonino Marcian\`o}
\email{marciano@fudan.edu.cn}
\affiliation{Center for Field Theory and Particle Physics \& Department of Physics, Fudan University, 200433 Shanghai, China}

\author{Yong-Shi Wu}
\email{yswu@fudan.edu.cn}
\affiliation{Center for Field Theory and Particle Physics \& Department of Physics, Fudan University, 200433 Shanghai, China}
\affiliation{
Department of Physics and Astronomy, University of Utah, Salt Lake City, Utah, 84112, U.S.A.} 

\begin{abstract}
\noindent
Hawking, Perry and Strominger recently invoked BMS symmetry charges in an attempt to resolve the black hole information paradox. Here we propose an alternative scenario that is based on the Kac-Moody charges. We show that the role of BMS charges can be played by an infinite set of symmetries that emerge from the space-time foam predicted by quantum gravity. Specifically, we focus on Yang-Mills fields embedded in the gravity described by the Holst formulation, and argue that the Yang-Mills and gravitational self-duality conditions in space-time bubbles are related to a new infinite dimensional global symmetry, hidden in the Lagrangian. Such a symmetry is manifested by the Kac-Moody algebra, with zero central charges. This implies the existence, in the space-time foam, of an infinite number of different instantons that are interconnected by the Kac-Moody symmetry. These modes puncture the horizons of the building block of the space-time bubbles. On the other hand, the same Kac-Moody symmetry is retried in non-perturbative regime at the level of the gravitational quantum loops. The new result carries consequences on the no-hair theorem and on the study of quantum black holes. In particular, instantonic moduli of the Kac-Moody charges are quantum hairs encoding the missing black hole information, subtly compatible with the no hair theorem. 

\noindent 

 \end{abstract}
\maketitle

{\it{\textbf{Introduction.}}}
Global symmetries in gauge theories and embedded in the gravity have been intensively studied over the last few decades. Recently this attention underwent a lusty revival, due to the discovery that a new infinite dimensional $U(1)$ Kac-Moody algebra of symmetries \cite{Strominger:2013jfa,He:2014laa,Strominger:2014pwa,He:2014cra,He:2015zea,Dumitrescu:2015fej} emerges in the soft infrared limit of scattering amplitudes, without manifesting itself in the starting Lagrangian \footnote{Another attempt that involves an infinite dimensional algebra different than BMS was proposed in Refs.~\cite{Ellis:2016atb}.}. These are the BMS symmetries that arise asymptotically for a certain class of gravitational backgrounds, which are also investigated in connection with black hole physics and the resolution of the information paradox \cite{Hawking:2016msc,Dvali:2015rea,Ellis:2016atb}. Similar to Refs.~\cite{Strominger:2013jfa,He:2014laa,Strominger:2014pwa,He:2014cra,He:2015zea,Dumitrescu:2015fej}, but a few years in advance, few seminal papers \cite{Chau:1981gi,Hou:1981hn,Chau:1982mn,Tze:1982gf,Ivanova:1997cu,Popov:1998pc,Adam:2008jx} unveiled that non-local Kac-Moody symmetries can emerge by imposing the self-duality constraint in Yang-Mills theories defined on a complexified Euclidean space-time --- see \cite{KacMoody} for references on the Kac-Moody algebra. Intriguingly, both frameworks seem to provide counter-examples against the common ``Lagrangian paradigm": not all the symmetries of a field theory show themselves in the perturbative Lagrangian.

An inevitable implication of semiclassical quantum gravity is the emergence of a space-time foam at length scales close to the Planck scale. For instance, the presence of virtual black holes is implied by the Heisenberg's uncertainty principle. These can be perceived as a basis for the space-time foam. According to Hawking, Page and Pope \cite{HPP}, the space-time foam can be topologically deconstructed into three building blocks, $S_{2}\times S_{2}, K_{3}, CP_{2}$, dubbed {\it gravitational bubbles}. Contrary to gravitational instantons, gravitational bubbles are not the solution of the (Euclidean) Einstein's field equations. They are instead quantum fluctuations of the space-time geometry. We refer to them as happening at ``mesoscopic scales'', at which space-time is not purely classical nor purely quantum.

We propose here a solution to address the fate of the black hole (BH) information loss paradox that radically differs from other studies based on the BMS symmetry argument --- these were developed in recent years to address first the case of hypercharge fields \cite{Strominger:2013jfa,He:2014laa,Strominger:2014pwa,He:2014cra,He:2015zea,Dumitrescu:2015fej}, then the case of YM fields and the gravity \cite{Hawking:2016sgy}. Although both the BMS symmetry inspired argument and ours are based on new infinite dimensional symmetries --- specifically, in our case we deal with a Kac-Moody algebra without central charge term --- the mechanism of quantum information processing that we hypothesize is totally different. As recently pointed out in \cite{Bousso:2017dny} by recovering canonical transformations that decouple the soft variables from the hard dynamics, large gauge symmetries (such as BMS ones) are not relevant to the fate of the BH information paradox --- long-wavelength photons or gravitons undergo only trivial scattering in their own sector, simply passing through the interaction region. Further criticism concerning quantum encoding was summarized in \cite{Donnelly:2017jcd} by pointing out that for theories deploying the BMS argument, localized information is independent of fields outside a region, hence implying that ``soft hairÓ play no role in encoding information. As an alternative to this scenario, we do not move from any argument involving photon hairs and their scattering. We suggest instead that quantum hairs may be identified as excited moduli of instantonic modes that can be recovered on the foamy texture of space-time, at mesoscopic scales. 

These reasonings motivate us to make contact with prior studies in the '80s \cite{Chau:1981gi,Hou:1981hn,Chau:1982mn,Tze:1982gf,Ivanova:1997cu,Popov:1998pc,Adam:2008jx}, reanalyze the self-duality condition, which leads to gauge instantons in \emph{4D} space-time, and relate it to the Kac-Moody symmetry. We specify the link between the self-dual Yang-Mills theories on complexified flat spacetimes and the YM theories embedded in the space-time foam. We find that, contrary to YM theories defined in a Minkowksi space-time, in the case of the space-time foam there exist an infinity of different YM instantons with the same standard moduli interconnected by an infinite dimensional Kac-Moody algebra. We show how to extend the same result to gravitational instantons.

We then focus on the infinite dimensional Kac-Moody algebra of symmetries. In light of previous attempts (for most recent notable proposals see Refs.~\cite{Almheiri:2012rt, Almheiri:2013hfa,Maldacena:2013xja}), the argument based on Kac-Moody symmetries represents a new breakthrough. According to the argument based on the BMS symmetries (and the BMS charges), information is supposed to be stored not in the interior of the black hole, but on its event horizon \cite{Hawking:2015qqa}. As we will see, charges associated with Kac-Moody self-dual instantons on space-time bubbles give rise to a more radical picture. Zooming in to smaller (mesoscopic) scales, we can unveil the building blocks of the space-time foaminess, which are associated with semiclassical quantum gravity. This scenario suggests that information is stored not only on the boundary of the black hole, but everywhere in the region surrounding the (would-be) singularity, both inside and outside the event horizon. Such a picture implements the original idea of holography --- as stated in Ref.~\cite{tHooft:1993dmi,Susskind:1994vu} and further developed in the framework of string theory (see e.g. Refs.~\cite{Maldacena:1997re, Maldacena:2013xja}) --- only at mesoscopic scales, on the virtual building blocks of the space-time foam. Conversely, holographic principle ceases to be valid at macroscopic scales according to our theory.

The methodology we invoke in this paper makes use of a mesoscopic analysis of the instantonic YM and gravitational solutions on the space-time foaminess predicted by quantum gravity at the semiclassical level. We further investigate the (microscopic) Planckian structures of the underlying fundamental objects appearing in our construction, through the lenses of a model of quantum gravity,  loop quantum gravity (LQG) \cite{Rovelli:2004tv}, which realizes the diffeomorphism invariant quantization of the relevant self-dual variables smeared on a lattice. 

Respectively, at mesoscopic scales YM instantonic solutions are recovered on the building blocks of the space-time foaminess, which include $S_{2}\times S_{2}$ topologies --- these are virtual black hole pairs fluctuations of the geometry. Since instantonic solutions come with an infinite amount, and are interconnected by the infinite dimensional Kac-Moody symmetry, they can be claimed to carry an infinite number of hairs, which are of quantum nature. This suggests a way out to the no hair theorem \cite{Veneziano:2012yj, Coleman:1991ku}, and consequently to the black hole information paradox. Information is stored in the quantum hairs, while quantum corrections may introduce mass-gaps in the model, reducing the degeneracy of the information that can be stored in the instantonic levels \footnote{Classical moduli of instantons in the $S_2\times S_2$ geometry are gapless. Their number is infinite and corresponds to all the Kac-Moody generators of the symmetry. Generically, the quantum moduli space of an instanton is not the same of the classical one --- for example, this can be explicitly shown in $\mathcal{N}=1$ SUSY YM models. In a realistic model, one has also to consider interactions among instantons beyond the instantonic gas approximation. This may reduce the gapless moduli space: a finite large number of moduli can then acquire a mass-gap. The information from the gravitational collapse can be processed in mass-gap levels of a finite large number of moduli.}. 

Finally, we show that a Kac-Moody symmetry is retried in the non-perturbative quantum gravity regime as a symmetry of the gravitational Wilson loops. In particular, within the framework of LQG we will argue how {\it different gravitational Wilson Loops are connected by a Kac-Moody symmetry} very much in the same way as that of instantons.  Thus such a symmetry can be thought as an infinite dimensional symmetry for a subset of states within the kinematical Hilbert space of LQG.

As a consequence, such an infinite amount of quantum moduli is obtained, which may be reinterpreted as {\it quantum black hole hairs}.  As suggested by {\it Veneziano \cite{Veneziano:2012yj}, Coleman, Preskill and Wilczek} \cite{Coleman:1991ku}, quantum hairs are subtly compatible with the classical {\it no hairs theorem} and they may provide the missing contribution to the black hole information, i.e. they may be the key to solve the black hole information paradox. Our conjecture is that quantum hairs are originated from instantonic moduli at semiclassical level. In particular, infalling collapsing matter into the black hole excites ground state instantons of determined size and center into instantons with the same size and center but different (quantum reduced) Kac-Moody charges. We give a heuristic insight on this perspective in Fig.~1. 
Let us remark that, even if the number of moduli is infinite, only a finite number of them will be excited, rendering a finite contribution to the BH entropy. This is because the infalling information during the BH collapse is finite, and thus can only excite a finite number of moduli. We will also argument how interactions of instanton dynamically breaks this symmetry, i.e. the Kac-Moody current gets an anomalous contribution from quantum corrections. As a result, zero mode fields, associated to moduli, will become pseudo-goldstone bosons of the Kac-Moody symmetry. This is also necessary to generate a mass gap of moduli.
\\

\begin{figure}[t]
\centerline{ \includegraphics 
[scale=0.06]{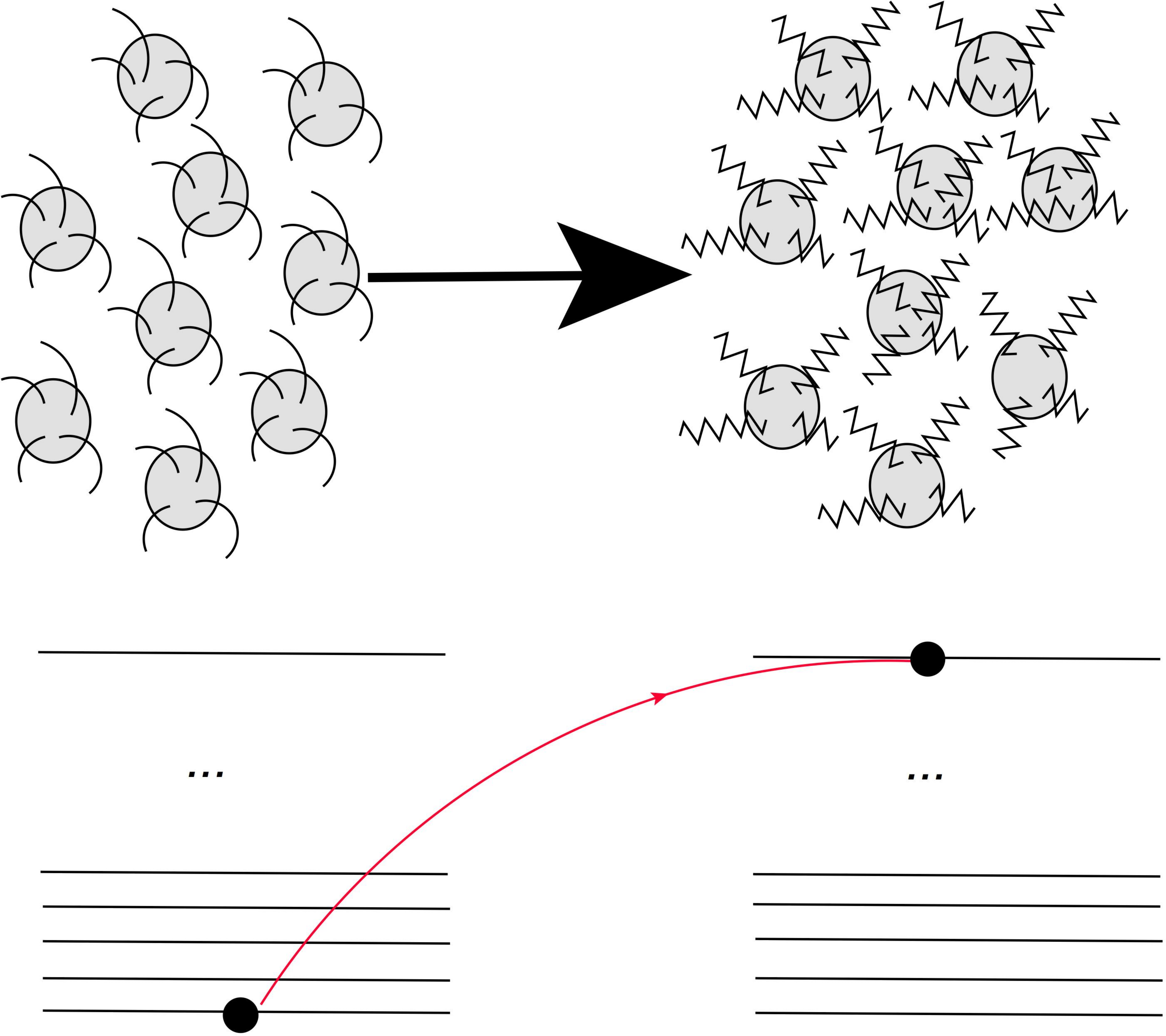}}
\vspace*{-1ex}
\caption{A system of instantons with Kac-Moody moduli pictorially represented as hairs.
The black hole change of state corresponds to a transformation of the instantonic hairs.
This changes the energy state of the system and the information processed 
through mass gap of energy levels. }
\label{plot}   
\end{figure}

{\it{\textbf{Bubbles and Kac-Moody algebra.}}}
Here we develop a different argument for the resolution of the information paradox, which is related to the emergence of an infinite dimensional Kac-Moody algebra of symmetries, and its subsequent regularization due to the quantum gravitational area-gap. We consider the propagation of YM fields on space-time bubbles, and focus on $S_{2}\times S_{2}$ topologies. According to Hawking, Page and Pope, these may be physically interpreted as virtual black hole pairs --- other gravitational bubbles like $K_{3}$ only provide sub-leading corrections to the space-time foam.
The topology may be constructed from $S_{4}$-sphere through a series of simple operations. The $S_{4}$-sphere is equivalent to a $\mathbb{R}^{4}$-space completed with a point $a$ added at infinity. On the $S_{4}$-sphere we can fix and identify two specular points $b$ and $c$, with angular coordinates $(\theta,\phi\,\chi,\psi)$ and $(\pi-\theta,-\phi,\pi-\chi,\pi-\psi)$ at the equator $\psi=\pi/2$. This operation corresponds to an orbifold $S_{4}/\mathbb{Z}_{2}$. Now, we cut off the neighborhoods of $b$ and $c$, replacing them respectively with an Eguchi-Hanson and an Anti-Eguchi-Hanson metric. We can now consider a point $a$ in the neighborhood of $b$ (or also in the neighborhood of $c$, given the symmetry of the problem) and send it to infinity through a conformal transformation. In this way we obtain an asymptotically Euclidean metric, with topology $S_{2}\times S_{2}-\left\{p\right\}$. In the literature, this construction is shortly dubbed as $S_{2}\times S_{2}$ bubble. The $S_{2}\times S_{2}$ bubble may topologically correspond to the gravitational instanton 
with an Euclidean Nariai metric
 \be\label{Nariai}
ds^{2}=\left(d\Omega_{(2)}^{2}(\psi,\chi)+d\Omega_{(2)}^{2}(\theta,\phi)\right)\,,
\ee
where radii of each $S_{2}$ 2-spheres are normalized to be one
and 
\begin{eqnarray}
&&d\Omega_{(2)}^{2}(\theta,\phi)=d\theta^{2}+\sin^{2}\theta d\phi^{2}\,,\nonumber \\
&&d\Omega_{(2)}^{2}(\psi,\chi)=d\psi^{2}+\sin^{2}\psi d\chi^{2}\,. \nonumber
\end{eqnarray}
for $0 \leq \theta <\pi$, $0 \leq \theta <2\pi$,  $0 \leq \psi <\pi$, $0 \leq \chi <2\pi$. 

This metric can be recast 
by means of the coordinates 
\begin{eqnarray}
&&x_{1}=r_{1}\cos \phi,\,\,\,x_{2}=r_{1} \sin \phi\,,\nonumber \\
&&x_{3}=r_{2} \cos \chi,\,\,\,x_{4}=r_{2} \sin \chi\,,\nonumber \\
&&r_{1}=\sqrt{x_{1}^{2}+x_{2}^{2}},\,\,\,r_{2}=\sqrt{x_{3}^{2}+x_{4}^{2}}\,, \nonumber 
\end{eqnarray}
as 
 \be\label{Nariai2}
ds^{2}=4\left(\frac{dx_{1}^{2}+dx_{2}^{2}}{(1+r_{1}^{2})^{2}}+\frac{dx_{3}^{2}+dx_{4}^{2}}{(1+r_{2}^{2})^{2}}\right)\,.
\ee

The same metric can be finally complexified, using 
\begin{eqnarray}
&&z_{1}=(x_{1}+ix_{2}),\,\,\,\bar{z}_{1}=\left(x_{1}-ix_{2}\right)\,, \nonumber \\
&&z_{2}=\left(x_{3}+ix_{4}\right),\,\,\,\bar{z}_{2}=\left(x_{3}-ix_{4}\right)\,, \nonumber
\end{eqnarray}
to
 \be\label{Nariai3}
ds^{2}=\left(\frac{dz_{1}d\bar{z}_{1}}{(1+|z_{1}|^{2})^{2}}+\frac{dz_{2}d\bar{z}_{2}}{(1+|z_{2}|^{2})^{2}}\right)\,.\ee
This is equivalent to state that $S_{2}\times S_{2}\simeq CP_{1}\times CP_{1}$, where $CP_{1}$ manifolds have Fubini-Study metrics. In other words, the two $S_{2}$-spheres are mapped into two Riemann spheres. Our following analysis is performed assuming the two Riemann spheres radii to be constant.

On this line element, outside the mathematical singular point connecting the two Riemann spheres, we consider a $SU(N)$ YM theory with the self-duality constraint 
 \be\label{Spi}
F_{\alpha\beta}=\tilde{F}_{\alpha \beta}\,,
\ee
in which $\tilde{F}_{\alpha\beta}=\frac{1}{2}\epsilon_{\alpha\beta}^{\ \ \ \gamma \delta}F_{\gamma \delta}$. 

The self-dual equations in complex coordinates recast
 \be \label{correspond}
 F_{z_{1}z_{2}}=F_{\bar{z}_{1}\bar{z}_{2}}=F_{z_{1}\bar{z}_{1}}+F_{z_{2}\bar{z}_{2}}=0\,.
 \ee
We can now follow the ADHM construction proposed in \cite{ADHM}, adapted to the $S_{2}\times S_{2}$ background. Eq.~(\ref{correspond}) implies that the gauge field components can be expressed in terms of $(N+2K)\times N$ complex matrices $\mathcal{D}_i$ (with complex conjugate matrices $\bar{\mathcal{D}}_i$) --- here $K$ stands for the instantonic topological charge and $i=1,2$ --- such that
 \begin{eqnarray} 
&& A_{z_{1}}=\mathcal{D}_{(1)}^T\mathcal{D}_{(2),z_{1}},\,\,\,\,A_{z_{2}}=\mathcal{D}_{(1)}^T\mathcal{D}_{(2),z_{2}}  \label{Axx}
\\
&& A_{\bar{z}_{1}}=\bar{\mathcal{D}}_{(1)}^T\bar{\mathcal{D}}_{(2),\bar{z}_{1}},\,\,\,\,A_{\bar{z}_{2}}=\bar{\mathcal{D}}_{(1)}^T\bar{\mathcal{D}}_{(2),\bar{z}_{2}}\,,  \label{Axxbar} 
 \end{eqnarray}
 where $\mathcal{D}_{(i),z_{1}}=\partial_{z_{1}}\mathcal{D}_{(i)}$ and so forth. Finally, we define the matrix $\mathcal{J}$ to be
 \be \label{J}
 \mathcal{J}=\mathcal{D}^T_{(2)}\mathcal{\bar{D}}_{(1)}\,.
 \ee
The matrices $\mathcal{D}_{(i)},\mathcal{\bar{D}}_{(i)}$ have the geometrical interpretation of phase rotation factors in the complex two-dimensional space of $(z_{1},\bar{z}_{1})$ and $(z_{2},\bar{z}_{2})$. This is a very important difference with respect to the standard ADHM construction: $\mathcal{D}^T_{(i)}$ and $\mathcal{\bar{D}}_{(i)}$ are not hermitian conjugates of each other, as opposed to that for the simply connected Minkowski space-time.

We can rewrite the field strength as 
\be \label{Fuv}
F_{\zeta,\bar{\zeta}}=-\overline{\mathcal{D}}_{(1)}(\mathcal{J}^{-1}\mathcal{J}_{,\zeta})_{,\bar{\zeta}}\overline{\mathcal{D}}_{(2)},\,\,\,\,\,\,(\zeta=z_{1},z_{2})\,
\ee
The self-dual constraint (\ref{Spi}) corresponds to having the constraint on the $\mathcal{J}$-components as 
\be \label{JJyybar}
(\mathcal{J}^{-1}\mathcal{J}_{,z_{1}})_{,\bar{z_{1}}}+(\mathcal{J}^{-1}\mathcal{J}_{,z_{2}})_{,\bar{z}_{2}}=0\,.
\ee
The gauge transformations are
\be \label{Gauge}
A_{\mu}\rightarrow \mathcal{G}^{-1}A_{\mu}\mathcal{G}+\mathcal{G}^{-1}\mathcal{G}_{,\mu}\,,
\ee
which correspond to 
\be \label{DDbarg}
\mathcal{D}^T_{(1)}\rightarrow \mathcal{G}^{-1}\mathcal{D}^T_{(1)},\,\,\,\mathcal{\overline{D}}_{(1)}^{T}\rightarrow \mathcal{G}^{-1}\mathcal{\overline{D}}_{(1)}^{T}\,,
\ee
\be \label{DDbarg2}
\mathcal{D}_{(2)}\rightarrow \mathcal{G}\mathcal{D}_{(2)},\,\,\,\mathcal{\overline{D}}_{(2)}\rightarrow \mathcal{G}\mathcal{\overline{D}}_{(2)}\,,
\ee
in which $\mathcal{G}$ is a $SL(N,\mathbb{C})$ matrix. 

Because of the initial self-duality condition, the $\mathcal{J}$-current respects a global infinite dimensional algebra \cite{Chau:1981gi,Hou:1981hn,Chau:1982mn,Tze:1982gf}
\be \label{algebra}
[\delta_{\alpha}^{(m)},\delta_{\beta}^{(n)}]\mathcal{J}=\alpha^{a}\beta^{b}C_{ab}^{c}\delta_{c}^{(m+n)}\mathcal{J}\,,
\ee
in which $\alpha^a,\beta^a$ are infinitesimal shift parameters, while $C^a_{\,bc}$ are group structure  matrices. Brackets in Eq.\eqref{algebra} define a Kac-Moody algebra \cite{KacMoody}. $\mathcal{J}$ turns out to be the so called Kac-Moody current: it is shifted by the Kac-Moody group generators and its shift respects the Kac-Moody algebra. The latter corresponds to 
\be \label{algebra}
[\mathcal{Q}_{a}^{(m)},\mathcal{Q}_{b}^{(n)}]=C_{ab}^{c}\mathcal{Q}_{c}^{(m+n)}\,,
\ee
in which
\be \label{charge}
\mathcal{Q}_{a}^{m}=-\int d^{2}z_{1} d^{2}z_{2} {\bf Tr}\left[\delta_{a}^{(m)}\mathcal{J}\frac{\delta }{\delta \mathcal{J}} \right]
\ee
are the conserved Noether charges. 
We emphasize that information on the topological charge $K$ is contained in the current $\mathcal{J}$, as a $(N+K)\times (N+K)$ matrix combining two $N\times (N+2K)$ matrices. For every $K$-charge, the algebra structure is unchanged. This is not only a formal mathematical result. It implies that the Kac-Moody algebra does not change topological charge, i.e. Kac-Moody charges commute with topological charges. 

Appropriate reality conditions must be satisfied. Specifically, $A_{\mu}^{a},\,\mathcal{J}$ must be positive definite. Another requirement is that the variations of  $\mathcal{J},\bar{\mathcal{J}}$ are Hermitian: $\delta \mathcal{J}=(\delta \mathcal{J})^{\dagger}$.

Because of Eq.~(\ref{Fuv}), when $\mathcal{J}$ shifts the field strength of the instanton shifts consequently as 
\be
F\rightarrow F+\delta F\,,
\ee
with 
\be
\delta F=-\overline{\mathcal{D}}_{(1)}^T(\delta \mathcal{J}^{-1}\,\mathcal{J}_{,\zeta}) \overline{\mathcal{D}}_{(2)}-\overline{\mathcal{D}}_{(1)}^T( \mathcal{J}^{-1}\,\delta \mathcal{J}_{,\zeta}) \overline{\mathcal{D}}_{(2)}\,.
\ee
This shift is not just a gauge artifact, but is related to the shift of the Kac-Moody current. The kinetic term in the Lagrangian does not shift under gauge transformation, but the $\mathcal{J}$-current does. This ensures that the instantonic solutions are not invariant under the Kac-Moody symmetry, and shift with the shift of $\mathcal{J}$. Thus within this YM scenario Kac-Moody symmetries connect an infinite number of different instantonic solutions, which have exactly the same shape and center but not the same Kac-Moody charges. The shift is at discrete levels, following the Kac-Moody descent. \\

{\it{\textbf{Kac-Moody symmetries for 
gravity.} }}
It is remarkable that the Kac-Moody algebra can be recovered also from the action of Einsteinian gravity. The key point to start with is that the Einstein-Hilbert action can be cast in term of SU(2) gauge-variables, which in the original formulation of Ref.~\cite{Ashtekar} are self-dual connections. Denoting the gravitational self-dual connection as $A$ --- and its field strength as $F=dA+ A\wedge A$, in which contraction of internal indices involve the structure constants of SU(2) --- and introducing a soldering $1$-form $\gamma$ such that the metric tensor with Lorentzian signature reads $g_{\mu \nu}={\bf Tr}\, \gamma_\mu^\dagger \gamma_\nu$ and the Plebanski $2$-form $\Sigma= \imath \, \gamma^\dagger \! \wedge \gamma$ can be defined, the action for gravity reads 
\be\label{eh}
S_{\rm EH}=  \kappa \int {\bf Tr}\,\, \Sigma\wedge F- \frac{\lambda}{2}\,  {\bf Tr}\,\, \Sigma\wedge \Sigma\,,
\ee
where $\kappa=(8\pi G_N)^{-1}$ and $\lambda$ denotes the cosmological constant. The equations of motion are the Gau\ss\, constraint $d_A\wedge \Sigma=0$ --- here $d_A$ denotes the covariant derivative with respect to the self-dual gravitational connection $A$ and the constraint generates internal SU(2) transformations --- that encodes the Cartan structure equation, and 
\be \label{gd}
\gamma\wedge F=\lambda\, \gamma \wedge \Sigma \,.
\ee
The similarity between the role of $\Sigma$ in \eqref{eh} and the role of $\tilde{F}$ in the Yang-Mills action --- we can cast the latter as $\int F\wedge \star F$ in terms  of $\star$, the internal Hodge operator \footnote{On-shell the actions of the internal and external Hodge dual operators coincide in a space-time with $4$D Euclidean signature.} --- suggests that \eqref{gd} can be solved assuming the {\it ansatz} $F=\lambda \Sigma$, analogous to $F=\star F$. Using this {\it ansatz} and taking into account the Bianchi identity $d_A\wedge F=0$, it immediately follows \cite{Samuel:1988jx} that gravitational instantonic solutions can be recovered --- notice that not only the connection but also the Plebanski two form is self-dual.

It is also possible to show (see e.g. Ref.~\cite{Oh:2011nv}) that Yang-Mills instantonic solutions can be always found on a background that is itself a gravitational instantons. Namely, one can show that all the gravitational instantons are SU(2) Yang-Mills instantons. 
Since the analysis of gravitational instantons was performed on a principle SU(2) bundle, we can easily extend the procedure outlined in \cite{Oh:2011nv}, and cast the self-dual equations for the gravitational field in complex coordinates, as in \eqref{correspond}. This allows to define for the gravitational case a Kac-Moody current $\mathcal{J}$, on the same foot as in Eqs.~\eqref{Axx}--\eqref{J} for Yang-Mills SU(N) gauge fields. The infinite dimensional Kac-Moody algebra introduced in Eq.~\eqref{algebra} immediately follows for the gravitational instantons.

\begin{figure}[t]
\centerline{ \includegraphics [scale=0.10]
{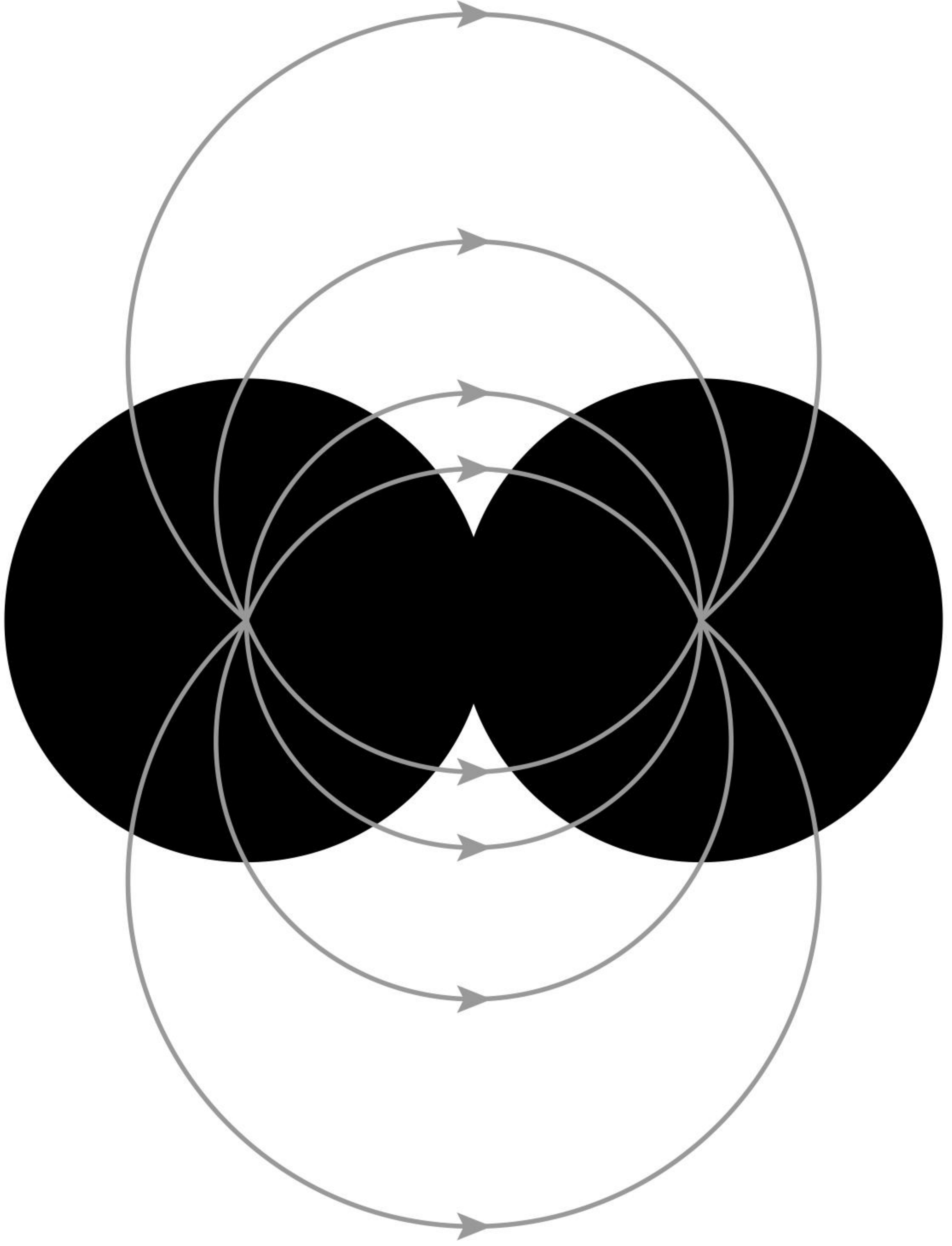}}
\vspace*{-1ex}
\caption{Punctures on $S_{2}\times S_{2}$ are displayed in gray. They are associated with
further extra contributions from gravitational instantons in the semiclassical approximation, 
while in the non-perturbative regime they correspond to gravitational Wilson lines. }
\label{plot}   
\end{figure}

\vspace{1cm}
{\it{\textbf{Isolated horizons and Chern-Simons theory.}}}
Gravitational instantonic solutions such as the Eguchi-Hanson metric can be recast in term of self-dual SU(2) Ashtekar variables, as clarified in Ref.~\cite{Oh:2011nv}. This opens a clear pathway to shed light on instantonic solutions from the perspective of isolated horizons (IH), introduced in Ref. \cite{Ashtekar:1999wa}. These are non-expanding horizons (NEH), which are defined for a spacetime on manifold $\mathcal{M}$ and with metric $g_{ab}$ provided that: {\it i)} they represent a spacetime boundary that is topologically $S_2\times \mathbb{R}$;  {\it ii)} given any null normal $\ell^a$ and $q_{ab}$, the pull-back of the space time metric, the expansion rate $\theta_{(\ell)}=q^{ab}\nabla_a\ell_b$ is vanishing; {\it iii)} 
all field equations hold on the NEH, and for the pull-back of the energy-momentum tensor it holds that $-T^a_b \ell_b$ is causal and future oriented for any future directed null normal $\ell^a$. Conditions {\it i)} and {\it iii)} imply that, given the geodesic normal field $\ell_a$, for the surface gravity $\kappa_{(\ell)}$ it holds that $\ell^b\nabla_b \ell^a=\kappa_{(\ell)} \ell^a$, where $\nabla_a$ is a covariant derivative compatible with $g_{ab}$.  Then the IH can be introduced as a NEH equipped with an equivalence class of null normals that satisfies the requirement $[\mathcal{L}_{\ell}, D_b]=0$, where $\mathcal{L}_{\ell}=\ell^a\nabla_a$ denotes the Lie derivative along $\ell^a$ and $D_b$ the intrinsic derivative operator, which is unique on NEH \cite{Ashtekar:2001jb}. From an effective point of view, IH look like ``apparent horizons in equilibrium'', and can generally account for rotation and distortion, their main property remaining that the surface gravity $\kappa_{(\ell)}$ is not evolving in time. 

A series of results --- see e.g. Refs.~\cite{Engle:2009vc, Engle:2010kt} --- allows to establish a link between the boundary of the Holst action \eqref{eh} on the IH and topological SU(2) Chern-Simons theory. We start considering the case of real Ashtekar-Barbero variables $A^i_\gamma$. We recast the Plebanski two form in \eqref{eh} in terms of tetrad fields that transform under the adjoint representation of $\mathfrak{sl}$(2,$\mathbb{C}$). The theory is now defined over a principle SL(2,$\mathbb{C}$) bundle, the Lorentz connection being labelled by a pair of antisymmetrized indices $I,J=1,..4$. The Plebanski two form then reads 
\be\nonumber
\Sigma^{IJ}=\left(\epsilon^{IJ}_{\ \ KL} +\frac{1}{2\gamma} \delta^{IJ}_{\ \ KL}  \right)\, e^K \wedge e^L\, ,
\ee
having denoted the Immirzi parameter with $\gamma\in \mathbb{R}$. On the boundary $\partial\mathcal{M}$, the Holst action provides the symplectic potential  
\be
\delta \mathcal{S}=\frac{2}{\kappa}\int_{\partial\mathcal{M}} \Sigma_i \wedge \delta K^i, 
\ee
with $K^i =A^{0i}$, from which the symplectic two-form immediately follows  
\be
\Omega(\delta_1, \delta_2)=\frac{1}{\kappa} \int_{\partial\mathcal{M}}  \delta_{1} \Sigma \wedge \delta_{2} K\,.
\ee
Denoting $\Gamma^i=-\frac{1}{2}\epsilon_{ijk}A^{jk}$ and setting the time gauge $e^0_i=0$, we can project the Cartan equations on $\partial\mathcal{M}$ and then find $\Sigma_i\wedge\delta\Gamma^i=-d(e^i\wedge \delta e_i)$. If we choose now $\partial\mathcal{M}$ to be an IH, denote the $2+1$ dimensional horizon with $\Delta$ and the two dimensional spatial section of the horizon with $H=\Delta\cap\mathcal{M}$, and introduce the Ashtekar-Barbero variables $A_\gamma^i=\Gamma^i + \gamma K^i$, we can show that the symplectic two form inherits a boundary term over $H$
\be
\gamma \kappa \Omega(\delta_1, \delta_2)=\int_\Delta \delta_{1}\Sigma_{i}\wedge\delta_{2} A^i + \int_H \delta_{1}e_{i}\wedge \delta_{2} e^i\,.
\ee  
On the sphere $H$, in virtue of the definition of IH, variations can be only expressed in terms of tangent diffeomorphism and SU(2) transformations. Calculating their contributions one is lead to (see e.g. \cite{Engle:2010kt})
\be
\int_H \delta_{1}e_{i}\wedge \delta_{2} e^i= -\frac{a_H}{2(1-\gamma^2)\pi} \int_H \delta_{1}A_{i}\wedge\delta_{2}  A^i\,,
\ee
where $a_H$ stands for the area of the sphere $H$, and the right hand-side represents the Chern-Simons symplectic two form for a real Ashtekar-Barbero connection restricted to the IH. The Ashtekar connection is now a dynamical Chern-Simons connection that satisfies the equation of motion 
\be \label{CSeom}
F^i_{ab}(A)=-\frac{2\pi}{a_H} \Sigma^i_{ab}\,.
\ee

{\it{\textbf{Microscopic loops.}}}
We can now take into account quantum states of the kinematical Hilbert space of the theory \cite{Rovelli:2004tv}. A graph $\Gamma$ is a collection of links $l$ intersecting at nodes $n$. It supports  spin-network states $|\Gamma, \{j_l, m_l\}, v_n \rangle $ that are labelled by spin $j_l$-representations of SU(2), their magnetic numbers $m_l$ and intertwines tensor representations $v_n$. Only the area operator will enter our considerations. We can suppress tensor representations and denote states as $|\Gamma, j_l, m_l \rangle $. The action of the right-hand-side of \eqref{CSeom} on these states reads 
\be 
\epsilon^{ab} \hat{\Sigma}^i_{ab} (x) |\Gamma, j_l, m_l \rangle = 2 \kappa \gamma \delta(x,x_p) \tau^i_p\, |\Gamma, j_l, m_l \rangle \,, 
\ee
$x_p$ denoting the coordinate where the link $l$ of $\Gamma$ pierces $H$, and $\tau^i$ a generator of the $\mathfrak{su}(2)$ algebra. This definition can be extended to the case of $p$ punctures on $H$, and then entails for Chern-Simons connections on the IH $\Delta$ the quantum dynamics 
\be \label{QEOM}
\hat{F}^i_{ab}(A)=-\frac{4\pi \kappa \gamma}{a_H} \sum_{p'=1}^{p} \delta (x,x_{p'})\,.
\ee

Chern-Simons action coupled to punctures on $H$ provides the theory to start from in order to derive \eqref{QEOM}, namely
\begin{eqnarray}
&\mathcal{S}= \mathcal{S}_{\rm CS} +\mathcal{S}_{\rm p}= \frac{k}{4\pi} \int_\Delta A^i \wedge d A_i +\frac{2}{3} A^i \wedge \epsilon_{ijk} A^j\wedge A^k \nonumber\\
&+ \Sigma_{p'=1}^{p}  \lambda_{p'} \int_{c_{p'}} {\bf Tr} 
[ \tau_3 \left( \Lambda_p\right)^{-1}\! \left(d_A \Lambda_p \right)] 
\,, \label{CSP}
\end{eqnarray} 
in which $\Lambda_p$ is a SU(2) group element that labels the position of the puncture, $c_p$ stands for the world-line of the puncture, $d_A$ is the covariant derivative with respect to the SU(2) connection, $\lambda_p$ is the coupling of punctures on $H$, and $\tau_{3}$ fixes an arbitrary direction in the internal space. \\
\\

{\it{\textbf{Gravitational entropy and instantons.}}}
Punctures have been thought so far \cite{Engle:2010kt} as defects on $H$ induced by degrees of freedom associated to point-like particles, whose positions on $H$ are labelled by SU(2) group elements and whose worldlines belong to $\Delta$. In our framework we rather think of punctures as the end-points of curves that have origin on another $2$ dimensions spatial surface $H$. Punctures' worldlines are then bijective maps between points of the topological 2-spheres that constitute the $S_2\times S_2$ building blocks of the space-time foam. For Euclidean YM SU(2) theories and SU(2) Holst gravitational theory, $SU(2)$ instantons (see e.g. BPST instantons in Ref.~\cite{BPST}) realize a mapping of 3-sphere into 3-spheres, and are then characterized by the homotopy classes $\pi_3(S_3)=\mathbb{Z}$. Wick rotating back to the Lorentzian metric the asymptotic 3-spheres $S_3$ become locally homeomorphic to $S_2\times \mathbb{R}$, restituting $2+1$ dimensional horizon. We might be tempted to interprete irreducible representations of SU(2) associated to punctures as the positive subset of $\mathbb{Z}$, thus labelling instantonic solutions that interconnects $\Delta$ hyper-surfaces boundaries of $\mathcal{M}$. Nonetheless gravitational instantonic solutions require the use of self-dual variables, with imaginary values of the Immirzi parameter. As we will show in a while, this leads to label representations with real numbers. 

Quantization of SU(2) Chern-Simons theories with real Ashtekar-Barbero connection encodes representations of  the quantum group SU\!\!\!$\phantom{a}_{\rm q}$(2), namely the deformation of the enveloping algebra of $\mathfrak{su}(2)$, often denoted as U\!\!\!$\phantom{a}_{\rm q}$(SU(2)). When the deformation parameter is the root of unity $q=e^{\imath \frac{\pi}{k+2}}$, the level $k$ of the Chern-Simons action naturally provides a regulator for the theory, the spin of the representations satisfying $j\leq k/2$. The dimension of the Hilbert space of the theory is then expressed by a regularized version of the Verlinde formula 
\be \label{Ver}
g_k(p,d_l)= \frac{2}{2+k} \sum_{d=1}^{k+1} \sin^2{\left( \frac{\pi d}{k+2} \right)} \prod_{l=1}^p \, \frac{\sin \left( \frac{\pi}{k+2} dd_l\right)}{\sin \left( \frac{\pi}{k+2} d\right) }\,,
\ee
in which $l$ labels the links that puncture. This expression is obviously finite, being the representations of the quantum groups bounded from above \cite{Geiller:2013pya}.

The case of self dual Ashtekar connection, which is of interest for us, can be obtained by an analytical continuation of \eqref{Ver} so to account for imaginary values of the Immirzi parameter. Specifically the analytical continuation reads \cite{Achour:2014eqa}
\be \nonumber
\gamma=\pm \imath \quad  \longrightarrow \quad  k=\mp \imath \lambda, \, \quad j_l=\frac{1}{2}(\imath s_l -1)\,,
\ee
with $\lambda, s_l\in \mathbb{R}^+$. This amount to a different choice of the gauge fixing while writing the Ashtekar variables, different than the temporal choice usually assumed \cite{Geiller:2011bh, Geiller:2012dd, Liu:2017bfk}. The internal gauge group of  symmetries for gravity becomes SU(1,1) \cite{Achour:2014eqa,Frodden:2012nu}. Analytical continuation thus provides a different spectrum of the area operator with respect to the one usually accounted for while dealing with discrete SU(2) representations \cite{Rovelli:2004tv}
\begin{eqnarray} 
&a_H(j_l)=8 \pi l_p^2\gamma \sum_l\sqrt{j_l(j_l+1)} \qquad\longrightarrow\nonumber\\
& a_H(s_l)=4 \pi l_p^2 \sum_l\sqrt{s_l^2+1}\,, \nonumber
\end{eqnarray} 
in which we set up $\gamma=\pm \imath$. It is relevant to notice that $s=0$ still provides an area gap. This implies that the level of the Chern-Simons action has a minimal value, since \cite{Engle:2009vc}
\be
a_H=4\pi \,l_P^2\, \lambda\,.
\ee
Such a procedure induces at the quantum level a mapping from representations of  U\!\!\!$\phantom{a}_{\rm q}$(SU(2)) to representations of  U\!\!\!$\phantom{a}_{\rm q}$(SU(1,1)). The Verlinde formula finally recasts as 
\be \label{Verc} \nonumber
g_\lambda=(n, s_l)=\frac{\imath}{\pi} \oint_\mathcal{C} dz  \sinh^2(z) \prod_{l=1}^{n} \frac{\sinh (\imath s_l z)}{\sinh z} \coth(\imath \lambda z)\,. 
\ee
The contour on the complex plane is $\mathcal{C}=\mathcal{C}_1\cup\mathcal{C}_2$, where $\mathcal{C}_1$ denotes any oriented path that lays in the first quadrant of the Argand-Gau\ss \, plane --- both $\mathfrak{Re}(z)\geq 0$ and $\mathfrak{Im}(z)\geq0$ --- with origin in $z=0$ and end-point in $z=\imath \pi$, and $\mathcal{C}_2$ denotes any oriented path that lays in the second quadrant --- $\mathfrak{Re}(z)\leq 0$ and $\mathfrak{Im}(z)\geq0$ --- starting in $z=\imath \pi$ and ending at $z=0$.  

The dimension of the kinematical Hilbert space is finite, since representations labelling its elements are bounded by the level of the Chern-Simons theory $\lambda$, the latter being connected to the area of the horizon by the formula $a_H=4\pi \,l_P^2\, \lambda$. In the semiclassical limit the dimension of the kinematical Hilbert space provides for the entropy to asymptotically approach the measure of $H$. Nonetheless, by assumption the building blocks we are considering are microscopic, and their entropies do not scale like the measure of $H$.  

The gravitational instantons, which at the semiclassical level are supported on continuum $S_3$ manifolds and are then Wick-rotated to $S_2\times \mathbb{R}$, can be thought at the quantum level to be defined in a distributional sense, on a finite set of worldlines that discretize the domain into a piece-wise linear manifold. \\ 
\\

{\it{\textbf{From loops to bubbles.}}}
Gravitational Kac-Moody symmetries were recovered at the semiclassical level, on space-time foam bubbles, to which we think now as quantum gravitational fluctuations of the metric within LQG. This allows us to use results on the measure of the Hilbert space to determine the entropy of the gravitational system. 

We seek to bridge between the two frameworks, and consider the space-time foam decomposition on $S_2\times S_2$, on which Eguchi-Hanson and Anti-Eguchi-Hanson metrics are picked. The latter are gravitational instantons cast in terms of self-dual Ashtekar variable \cite{Oh:2011nv}, representing  $2$ dimensional spatial sections $H=\Delta\cap\mathcal{M}$ of IH horizons $\Delta$. 

Kac-Moody instantons emerge when we consider the map between $\Delta$ sub-manifolds of BHs and anti-BHs, i.e. white holes (WHs). The map can be thought to be labelled by the subset of $s_l$ representations belonging to $\mathbb{N}$, which in turn corresponds to the homotopy class $\pi_3(S^3)$ of BPST instantons. Allowed value of the spin-representations are bounded by the level of the CS action on $\Delta$. The support itself of instantonic solutions, including punctures on $H$, is limited because for quantum fluctuations $a_H$ must be $O(l_P^2)$. Thus the volume of the Hilbert space is finite. 

Since space-time building block are far from the semiclassical asymptotic limit, the measure of the kinematical --- regularized because of the use of quantum representations --- Hilbert space is finite. The finiteness of the measure suggests that only a finite amount of instantons, with related Kac-Moody charges, must be considered when we focus on the quantum hairs related to the space-time foaminess. This suggests that information can be stored everywhere, inside or outside horizons, thanks a finite set of charges associated to the building blocks of the space-time foam.

All these arguments naturally lead to ask ourselves what happens to the Kac-Moody symmetry at the non-perturbative level, beyond semiclassical quantum gravity. The singular point in $S_{2}\times S_{2}$ induces a non-trivial Kac-Moody algebra also on the gravitational Wilson loops encircling the singular point. In particular, since the original Ashtekar connection satisfies the self-dual condition, an associated Kac-Moody algebra is expected. Shrinking the loop $U_{\gamma}$ towards the singular point allows the expansion 
\be \label{U}
U_{\gamma}=1\!\!1+\epsilon_{\gamma}^{2}\, \tau^{i}  F^{i}+ \cdots
\ee
where $F$ stands for the curvature of the Ashtekar connection and $\epsilon_{\gamma}$ denotes the side of an infinitesimal plaquette. $F$ can be then cast as in (\ref{Fuv}). Since as shown above curvature does not commute with the Kac-Moody symmetry, it is pretty evident from the expansion in (\ref{U}) that also the holonomy does not commute with the Kac-Moody algebra, i.e.
\be \nonumber 
{\rm Kac\!-\!Moody}\!:\,\, U_{\gamma}\rightarrow U_{\gamma}+\delta U_{\gamma}\, .
\ee
Thus we realize that {\it the Kac-Moody algebra connects an infinite number of gravitational loops associated to the same punctures}. Although gravitational Wilson lines cannot be identified with semiclassical gravitational instantons, nonetheless {\it the Kac-Moody symmetry remerges as a symmetry of the gravitational loops. }\\
\\

{\it{\textbf{Boundary terms and their effect.}}} The action of the YM theory on $S_{2}\times S_{2}$ can be recast by separating the $S_{2}\times S_{2}$ topology from the boundary surface connecting them, namely
\begin{eqnarray} \label{SSS}
&S=\int_{S_{2}\times S_{2}} d\zeta^{2} d\bar{\zeta}^{2}\frac{1}{4} {\bf Tr}\,F^{2}  \\
&=\int_{S_{2}\times S_{2}-B} d\zeta^{2} d\bar{\zeta}^{2}\frac{1}{4} {\bf Tr}\, F^{2}+\int_{B} d\mu \frac{1}{4} \, {\bf Tr}\, F^{2}\, , \nonumber
\end{eqnarray}
in which we suppressed internal and external indices, and denoted with $B$ the boundary surface separating the two $S_{2}$ spheres and with $d\mu$ the measure of coordinates adapted to $B$. The first contribution, on $S_{2}\times S_{2}-B$, can be cast as   
\begin{eqnarray} 
&\int_{S_{2}\times S_{2}-B} d\zeta^{2} d\bar{\zeta}^{2}\frac{1}{4} {\bf Tr}\,F^{2}= \nonumber\\
&= \int_{S_{2}\times S_{2}-B} d\zeta^{2} d\bar{\zeta}^{2}\left[\frac{1}{4} {\bf Tr}\,F\tilde{F}+\frac{1}{8} {\bf Tr}\,(F-\tilde{F})^{2}\right]  \nonumber\\
&=Q\frac{8\pi^{2}}{g^{2}}+\frac{1}{8}\int_{S_{2}\times S_{2}-B} d\zeta^{2} d\bar{\zeta}^{2} \,{\bf Tr}\,(F-\tilde{F})^{2}\, .
\end{eqnarray}
For the topological instantonic charge $Q>0$, the minimum of the action is obtained from the  instantonic condition $F=\tilde{F}$. Thus the contribution to the instantonic action reads 
\begin{eqnarray} 
S_{I}(\{S_{2}\times S_{2}-B\})=\frac{8\pi^{2}}{g^{2}}\, . \nonumber
\end{eqnarray}
However, another contribution to the instanton action is expected from the boundary manifold $B$, which is topologically equivalent to a point. In this region the instantonic solutions analyzed above seem not to be well defined. The self-duality condition $F_{\zeta,\bar{\zeta}}=\tilde{F}_{\zeta,\bar{\zeta}}$ was studied indeed only on the $S_{2}\times S_{2}-B$ background. Nonetheless, we may proceed by regularizing the entire action --- including the bulk action on the $S_{2}\times S_{2}-B$ manifold --- on infinitesimal Wilson loops $\alpha$ that realize a cellular complex decompositions of the manifold. The boundary action is then naturally captured by the regularization on the boundary Wilson loop $\alpha'$. Thus the whole action, regularized on a piece-wise linear distributional manifold, will be cast as
\begin{eqnarray} 
S\ =\!\!\!\!&&\sum_{\{\alpha\}} \frac{1}{4} \ {\bf Tr}\,\left[ U_{\alpha} \tau^A \right]\  {\bf Tr}\, \left[  U_{\alpha} \tau^A  \right] \nonumber\\
+\!\!\!\!&&  \frac{1}{4} \ {\bf Tr}\,\left[ U_{\alpha'} \tau^A \right]\  {\bf Tr}\, \left[  U_{\alpha'} \tau^A  \right]\,, \nonumber
\end{eqnarray}
in which $\tau^A$ denotes a generator of the internal algebra of the YM theory, $A$ stands for an index in the adjoint representation of the YM group and $\{\alpha\}$ represents the set of all the Wilson loops on the bulk.  Self-dual connections, to which correspond instantonic solutions, minimize both the bulk and boundary action, then introducing both Kac-Moody symmetries and charges on the manifold. As a remarkable non-local effect, the presence of a boundary, which we regularize as a Wilson loop, is also responsible for the generation of the Kac-Moody global symmetry of instantonic solution. \\

{\it{\textbf{Dynamical symmetry breaking of symmetries.}}}
We comment here on the dynamical symmetry breaking of Kac-Moody symmetries, and hence consider the action of an instanton and anti-instanton. In the limit in which their relative distance is assumed to be much higher than their size, the action can be divided into three terms, namely 
\begin{eqnarray}
S+S'+S_{\rm int}(x_{0},y_{0})
\end{eqnarray}
where 
\begin{eqnarray}
S=\int_{S_{2}\times S_{2}-B} d\zeta^{2} d\bar{\zeta}^{2}\left[\frac{1}{4}F\tilde{F}+\frac{1}{8}(F-\tilde{F})^{2}\right]\,, \nonumber\\
\end{eqnarray}
$S'=S[A']$ and the coordinates $x_{0}\equiv x_{0}(\rho, m_{i},A)$ are $y_{0}\equiv x_{0}(\rho', m_{i}',A')$ the instanton and anti-instanton centers, expressed as functionals of the two background connections $A$ and $A'$ of the two instantons. The action has no saddle point now for $F=\tilde{F}$. Instead it is minimized when 
\begin{eqnarray}
(F-\tilde{F})=-L_{int}\,.
\end{eqnarray}

The condition $F=\tilde{F}$ acquires an anomaly from quantum interactions among (anti)instantons. If $F\neq \tilde{F}$, the $J$ is no-more the Noether current of a Kac-Moody algebra.  Since the interaction terms introduce a continuos shift of the $F=\tilde{F}$ condition, not any non-trivial sub-algebra of the initial Kac-Moody symmetry is expected to be preserved. Thus we can easily conclude that the Kac-Moody algebra is {\it dynamically broken} by interactions among instantons. In the full non-perturbative regime, one should consider every infinite possible interactions among a large number of instantons and anti-instantons, which amounts to a complete loss of calculability. 

Instantonic moduli are associated to zero modes that are nothing but the Nambu-Goldstone bosons of the symmetry spontaneously broken by the instantonic solution itself. The fact that the Kac-Moody symmetry is dynamically broken means that the infinite number of Kac-Moody modes are not Nambu-Goldstone bosons anymore, but they rather gain dynamically a mass. In other words, they are pseudo-Nambu-Goldstone bosons. Moving form initial zero modes $a$, accounted as perturbations around the instantonic background, i.e. $A+a$, and considering the interaction terms, we then found that $a$ gets a mass gap. To exactly estimate their acquired masses seems to be impossible, as we should account for all the allowed interactions among instantons and anti-instantons in the many body limits. Nonetheless, we may still expect to recover a mass gap that is controlled by the confinement scale of the ($SU(N)$) YM theory taken into account. At the first level of the Kac-Moody ascendent scale $M\!=\!1$, we should then expect an energy level $E_{1}\!\sim\! \Lambda$, while at the $M$-th level an energy level $E_{M}\!\sim \!M\Lambda$, and so forth. As a result, $M$ different instantons connected by Kac-Moody transformations have an energy difference  $E_{M}-E_{M-1}\!\sim\! \Lambda$. \\

{\it{\textbf{Conclusions.}}}
We studied YM theories and the Holst action for gravity in the space-time foam background, focusing on the case of a $S_{2}\times S_{2}$ bubble, which is interpreted as a virtual black hole white hole pair. On this background, the self-duality condition, which is commonly associated to YM instantons, brings to a new (infinite dimensional) Kac-Moody algebra, with no central charge. Using self-dual gravitational connections, it is then possible to extend this construction to unveil the existence of a gravitational sector of the Kac-Moody algebra. 

This characterization is only vaguely reminiscent of the role of global BMS symmetries in QFT, which was recently investigated in a series of papers \cite{Strominger:2013jfa, He:2014laa, Strominger:2014pwa, He:2014cra, He:2015zea, Dumitrescu:2015fej}, and invoked in \cite{Hawking:2016msc} as a solution to the information paradox. Differently, here we found that {\it an infinite number of different Yang-Mills and gravitational instantons (with the same standard moduli) in space-time foam are connected by the Kac-Moody symmetry}. In other words, YM and gravitational instantons carry an infinite number of hairs. Nonetheless, we found a mass gap in this construction, which is mainly due to the discretization of space-time. 

Without entering the debate on the validity at the quantum level of the BMS symmetry approach developed in \cite{Hawking:2016msc}, {\rm which we emphasize in different from ours,} we rather point out that from our analysis Kac-Moody symmetries potentially emerge as playing a crucial role in the resolution of the information paradox. As in the case of the BMS symmetries, the infinite number of hairs is subtly compatible with the no-hair theorem and it may be connected to the solution of the black hole information puzzle. Differently than in the BMS picture, in our picture information is no more supposed to be stored at the event horizon, but everywhere around the (would-be) singularity. Bubbles that are topologically BH-WH pairs are the building blocks of the semiclassical limit, and to each generic element of the decomposition will correspond a family of gravitational instantons. The dimension of the Hilbert space and the associated entropy can be shown to be finite in this picture, being regulated by the level of the Chern-Simons theory on the boundaries $H$. The latter is in turn proportional to the measure of the horizon, and inherits an area gap that is proportional to the square of the Plank length $l_P$. Irreducible representations of punctures on $H$  label homotopy classes of instantons that interpolate between BH and WH 2-spheres constituting the $S_2\times S_2$ space-time bubbles. We can think at these instantonic solutions, and at the Kac-Moody charges that are hence associated, as virtual quantum hairs that are locally present, inside and outside BH horizons. It is worth noticing that the perspective we developed here, in which Kac-Moody charges are stored in the virtual BH pairs, is vaguely reminiscent of the ER$=$EPR conjecture \cite{Maldacena:2013xja}. But while the latter accounts for nontrivial stable gravitational configurations, in our approach information can be stored due to virtual processes of space-time fluctuations.\\

We finally point out that within the framework of quantum field theories in curved space-time, many other examples of symmetries that are hidden in the starting Lagrangians might be found, pointing towards the need of investigating the presence of new symmetries and checking the consistency of the proposals both at classical and quantum level in order to understand the information processing of black holes. \\
\\


\vspace{-0.5cm}

{\it{\textbf{Acknowledgement. }}}
We acknowledge enlightening discussions with A.~Ashtekar, J.~Ben Achour, M.~Bianchi, S.~Brahma, C.~Fields, J.~Lewandowski, C.~Rovelli and S.~Speziale.

\end{document}